\documentclass[article]{elsarticle}

\usepackage{booktabs} 
\usepackage{amssymb}
\usepackage{amsmath}
\usepackage{algpseudocode}
\usepackage{algorithm}
\usepackage{mathtools}

\usepackage[colorinlistoftodos,prependcaption,textsize=normalsize]{todonotes}
\usepackage[export]{adjustbox}
\usepackage{hyperref}
\usepackage{subfig}
\usepackage{multirow}
\usepackage{url}
\usepackage{textcomp}
\usepackage{xcolor}

\journal{Arxiv}

\begin{document}
\title{Viscovery: Trend Tracking in Opinion Forums based on Dynamic Topic Models}

\author[add1]{Ignacio Espinoza}
\ead{ignacio@novaviz.com}

\author[add1,add2]{Marcelo Mendoza\corref{cor1}}
\ead{marcelo.mendoza@usm.cl}
\cortext[cor1]{Corresponding author}

\author[add1]{Pablo Ortega}
\ead{pablo@novaviz.com}

\author[add1]{Daniel Rivera}
\ead{daniel@novaviz.com}

\author[add1]{Fernanda Weiss}
\ead{fernanda@novaviz.com}

\address[add1]{Universidad T\'ecnica Federico Santa Mar\'ia, Santiago, Chile}
\address[add2]{Centro Cient\'ifico y Tecnol\'ogico de Valpara\'iso, CCTVal, Chile}


\begin{abstract}
Opinions in forums and social networks are released by millions of people due to the increasing number of users that use Web 2.0 platforms to opine about brands and organizations. For enterprises or government agencies it is almost impossible to track what people say producing a gap between user needs/expectations and organizations actions. To bridge this gap we create Vis\-co\-ve\-ry, a platform for opinion summarization and trend tracking that is able to analyze a stream of opinions recovered from forums. To do this we use dynamic topic models, allowing to uncover the hidden structure of topics behind opinions, characterizing vocabulary dynamics. We extend dynamic topic models for incremental learning, a key aspect needed in Viscovery for model updating in near-real time. In addition, we include in Vis\-co\-ve\-ry sentiment analysis, allowing to separate positive/negative words for a specific topic at different levels of granularity. Vis\-co\-ve\-ry allows to visualize representative opinions and terms in each topic. At a coarse level of granularity, the dynamic of the topics can be analyzed using a 2D topic embedding, suggesting longitudinal topic merging or segmentation. In this paper we report our experience developing this platform, sharing lessons learned and opportunities that arise from the use of sentiment analysis and topic modeling in real world applications. 
\end{abstract}


\maketitle

\vspace{-2mm}
\section{Introduction}
\label{intro}

The emergence of the Web 2.0 has allowed that millions of users can send posts 
and opinions about celebrities, institutions, organizations and brands. As the volume of opinions in forums and blogs increases, the need to develop effective platforms for opinion search has become urgent. In the stream of opinions, trend tracking is a key building block of this kind of platforms, allowing to describe what users expect about institutions/organizations and how opinion trends evolve over time. 

Effective tools for opinion browsing need to incorporate opinion aggregation functionalities, being relevant to obtain descriptions of each trend. In addition, the sentiment orientation of opinions w.r.t. named entities lights up how users act/react in front of a given organization. Sentiment analysis methods are helpful in this task.

As the volume of opinions is huge, the need to develop effective aggregation me\-thods over opinions is the key building block of any opinion trend platform. Opinion clustering is a way to aggregate opinions. Using hard clustering algorithms each opinion can be assigned to a single class. However, documents achieve a best description by modeling its content with a mixture of topics, where each topic is defined as a probability distribution over words. In this way, opinions belong to several topics with different degrees of membership. This is the reason why documents are in general modeled using mixed membership models, and in particular Latent Dirichlet Allocation (LDA) models \cite{blei:03}, allowing to uncover the hidden structure of topics behind a corpus. LDA has made improvements in information retrieval tasks \cite{kurland:09} and outperforms standard text clustering algorithms being the state-of-the-art method for document aggregation.

In Chile, the National Agency of Consumers~\footnote{Sernac: Servicio Nacional del Consumidor, http://www.sernac.cl} centralize complaints about brands and their products. As it is almost impossible to follow each complaint, consumers may be disappointed due to the slow response of the Agency to their needs. To bridge this gap we created Viscovery, a platform for opinion aggregation and trend tracking that allow to browse a huge volume of opinions in a few minutes. The core of Viscovery is based on Dynamic Topic Models (DTM)~\cite{blei:06}, an extension of LDA~\cite{blei:03} capable of model a time sliced corpus, being able to estimate dependencies between vocabularies across time slices. To create Viscovery, we had to develop an incremental learning component able to update a model with new opinions. Our DTM update method achieves very similar results to DTM batch fitting in terms of topic coherence diminishing computational costs. In addition, we included in Viscovery sentiment analysis. Sentiment analysis allows to distinguish between subjective/neutral terms in each distribution of words, enlightening how consumers opine about brands and products. To include sentiment analysis in DTM we explore a simple approach based on aggregation, using lexical analysis at opinion level and conducting sentiment aggregations at topic and document level. A third element included in Viscovery is topic embedding. Using a time sliced 2D topic embedding, topic merging and topic segmentation are suggested. Dynamics across topics are very interesting for the analysis, and is a promising characteristic of Viscovery that allow practitioners to understand how topics evolve. Specific contributions of the paper are:

\begin{itemize}
\item A scalable implementation of DTM for online training updating model parameters when new opinions come to the platform.
\item A simple way to incorporate sentiment analysis into DTM, allowing to explore neutral/subjective words at different levels of aggregation.
\item A topic visualization tool that works with a time sliced 2D topic embedding, allowing to visualize how topics evolve over time.
\end{itemize}

The rest of the paper is organized as follows. In Section \ref{rel-work} we review related work on topic models and sentiment analysis. Incremental learning on DTM is presented in Section \ref{dtm_inc} and browsable sentiment analysis is discussed in Section \ref{browse_sent}. Section \ref{archi} presents the architecture of Viscovery. Implementation issues are discussed in Section \ref{imp_issues}. Viscovery data slices are presented in Section \ref{views} and finally we conclude in Section \ref{conc} giving conclusions and discussing future work. 

\section{Related Work}
\label{rel-work}

\paragraph*{Topic models} Main efforts on topic models start with probabilistic Latent Semantic Analysis \cite{hofmann:01} (pLSA), an aspect model for text developed using topic mixtures. This approach decomposes a corpus of documents across terms introducing latent variables, decoupling terms and documents with topic mixtures. Model fitting was conducted using the Expectation-Maximization algorithm (EM) \cite{dempster:77} casted for matrix completion with incomplete data. As the term-space is a high dimensional feature space, pLSA needs a high amount of data to perform well. As in general, text data is sparse, pLSA tends to overfit limiting generalization capabilities. To tackle this problem, Blei \textit{et al.} \cite{blei:03} introduced Dirichlet priors on vocabulary and document topic proportions. Using smoothing these models addressed the over-fitting limitations of pLSA. This kind of models, known as Latent Dirichlet Allocation (LDA) were firstly fitted using variational EM (VEM), an extension of the EM algorithm that successfully handle incomplete data with distributional priors. Later, Griffiths and Steyvers \cite{griffiths:04} explored Gibbs sampling for LDA model fitting, reducing the number of iterations until convergence. Gibbs sampling is the standard method used for LDA model fitting until today because its fast convergence does not affect the quality of the estimated models. Dynamic Topic Models (DTM) \cite{blei:06} was introduced to deal with vocabulary dynamics. DTM works over a corpus with timestamps, whilst model fitting is conducted using time slices of the corpus. Temporal dependencies across vocabularies are modeled using Kalman filtering, allowing to detect changes in descriptive words along different corpus slices. The inclusion of Kalman filtering in LDA for text dynamics involves additional computational costs in model fitting, slowing convergence. Despite computational costs involved in model fitting, DTM can successfully handle text dynamics.

\paragraph*{Sentiment analysis and topic modeling} A topic generative model for sentences with polarity was proposed by Eguchi and Lavrenko \cite{eguchi:06}. The model distinguishes between neutral words and sentiment words using a random binary variable that controls the membership of each word to each one of the  vocabularies. As documents can be generated from sentiment or topic words, each sentence achieves a polarity orientation calculated in terms of the number of sentiment words that contains. Dirichlet smoothing was used on topic and sentiment word distributions to avoid over-fitting. The performance of the model in information retrieval is tested inferring topic and sentiment orientation of each query showing that the proposal is feasible. Mei \textit{et al.} \cite{mei:07} proposed Topic Sentiment Mixture (TSM), a sentiment topic model with a two tier mixture of vocabularies to produce sentiment oriented sentences in a corpus. A first tier of the model is composed by neutral term distributions (one per each topic) and two additional term distributions for positive and negative words. Then, each topic can be produced by a mixture of these vocabularies defined from document proportions. The model is non-parametric (no distributional priors were used) and model fitting was conducted using the EM algorithm. TSM can be considered as an extension of pLSA to sentiment analysis being the main difference the split conducted over the vocabulary to distinguish between factual/subjective sentences. The Joint Sentiment Topic model (JST) based on LDA was proposed by Lin and He \cite{lin:09}. Term distributions were sampled over a simplex over terms cross polarities, then the generative model drawn topic proportions conditioned on each polarity. In this way, words can be drawn by topics $\times$ polarities distributions, producing words by the joint effect of topics and polarities in the document. As a consequence, sentiment coverages at document level can be directly estimated by the model. JST is able to successfully address the sentiment classification task at document level. An extension of JST was proposed by Jo and Oh \cite{jo:11}, who introduced Aspect and Sentiment Unification (ASUM). As JST, ASUM jointly models sentiments and topics, being topic proportions conditioned on polarities, with vocabularies at topic level per each sentiment orientation. However, ASUM models sentiment at sentence level, with words conditioned at a single topic per sentence. Results on sentiment classification shows that ASUM outperforms JST and comes close to supervised methods whilst ASUM does not require labels for model fitting. The state of the art shows that main efforts on sentiment topic modeling are focused on static models, discarding vocabulary dynamics. As the core of Viscovery is DTM, we will need to use a different approach to include sentiment analysis into dynamic topic models. We will show in Section \ref{browse_sent} how we use sentiment analysis at sentence level to conduct aggregation at different levels of granularities over DTM. 

\section{Incremental Learning for Dynamic Topic Models}
\label{dtm_inc}

\subsection{Dynamic Topic Models}

A set of latent variables can be introduced to model the relationships between terms and documents in a corpus. Formally, let $d \in \mathcal{D} = \{ d_1 , d_2 , \ldots , d_N \}$ and $w \in \mathcal{W} = \{ w_1, w_2, \ldots , w_M \}$ be random variables representing documents and terms, respectively. 
A set of random variables $z \in \mathcal{Z} = \{ z_1 , z_2 , \ldots , z_k \}$ can be introduced to model the joint probability of documents and terms, producing a mixed membership model expressed as follows:
\begin{equation} 
P(w | d) = \sum_{z \in Z} P(w | z) \cdot P(z | d). 
\label{eq:1}
\end{equation}
Using the Bayes rule to invert the conditional probability $P(z | d)$, we obtain an expression of the joint probability conditioned to the model parameters:
\vspace{-1mm}
\begin{equation} 
P(w,d) = \sum_{z \in Z} P(w | z) \cdot P(d | z) \cdot P(z).
\label{eq:2}
\end{equation}
The equation \ref{eq:2} is known as the generative formulation of the topic model of the corpus.

Topic models based on Dirichlet allocation require two Dirichlet distributions. 
A first one generates topic proportions for each document and a second one generates terms conditioned on document topics proportions.  
Specifically, a Dirichlet k-dimensional random variable $\theta$ takes values in a k-1 simplex ($0 \leq \theta_i \leq 1, \sum_{i=1}^k \theta_i = 1$), where its density function is defined by:

\begin{equation}
p(\theta | \alpha) = \frac{\Gamma (\sum_{i=1}^k \alpha_i)}{\prod_{i=1}^k \Gamma(\alpha_i) } \hspace{3mm} \theta_1^{\alpha_1} \cdot \ldots \cdot \theta_k^{\alpha_k},
\label{eq:5}
\end{equation}

\noindent and $\{ \alpha_1 , \ldots , \alpha_k \}$ corresponds to the distributional parameters, $\alpha_i > 0$. Then, equation \ref{eq:2} is expanded using Dirichlet priors:
\vspace{-1mm}
\begin{equation}
P(W,d) = \prod_{n=1}^{M} P(w_n | z_n , \beta) \cdot P(z_n | \theta_d) \cdot P(\theta_d | \alpha).
\label{eq:6}
\end{equation}

In equation \ref{eq:6}, $\theta_d$ indicates the proportion of topics in $d$. Then, $z_n$ is conditioned on $\beta$ and represents the 
sampling probability of $w_n$ on $d$. Note that $\alpha$ and $\beta$ are the distributional parameters of the Dirichlet density functions. 
Usually they are consigned as hyper-parameters to make a difference with model parameters. 
It is common to make an assumption of density symmetry for hyper-parameters, that is $\alpha_1 = \ldots = \alpha_k = \alpha$ and $\beta_1 = \ldots = \beta_k = \beta$. 
The values $\alpha$, $\beta$ control the level of smoothness/sharpness of the density functions around the centroid of the simplex. 

To model a time sliced corpus, Blei and Lafferty \cite{blei:06} introduced dynamic topic models (DTM). DTM is based on the static Latent Dirichlet Allocation model and use the mean parameterization of the multinomial topic distribution. The idea behind DTM is to use the mean parameterization of the topics to introduce mean chaining, being possible to model time dependencies over time. To chain topics over time, DTM models the chain of mean parameters introducing Gaussian noise, modeling uncertainty over time slices. Let $\beta_{t,k}$ be the $k$-th topic in the time slice $t$ and let $\pi$ be the mean parameter of the topic. Note that the $i$-th component of $\beta_{t,k}$ is given by $\beta_i = \log \left( \frac{\pi_i}{\pi_V} \right)$. As $\pi_i$ represents the expected value of $w_i$ and $\pi_V$ is the expected value of a random chosen word over the whole vocabulary $V$, the fraction $\frac{\pi_i}{\pi_V}$ is the odd of $w_i$ over $V$ and then $\beta_i$ corresponds to the \textit{logit} function for $w_i$ over $V$. As is known, a zero variation over $V$ achieves a zero value in the \textit{logit} function. Positive or negative deviations of $w_i$ in $V$ achieves positive or negative values in $[ -1 , +1 ]$, respectively. Then, $\beta_{t,k}$ can be chained in a state space of parameters that evolves with Gaussian noise: 
\vspace{-1mm}
\begin{equation}
\beta_{t,k} | \beta_{t-1,k} \sim \mathcal{N} (\beta_{t-1,k}, \sigma^2\mathcal{I}).
\label{eq:7}
\end{equation}

Topic proportions are also chained in DTM, using mean parameterization over $\theta$:
\vspace{-1mm}
\begin{equation}
\theta_t | \theta_{t-1} \sim \mathcal{N}(\alpha_{t-1}, \delta^2 \mathcal{I}).
\end{equation}

Time chaining does not affect model expressiveness. In fact, the decomposition of the joint distribution of words and documents in a corpus remains the same, except for the fact that both Dirichlet distributions (on topic proportions and terms) are conditioned on the Dirichlet distributions of the previous time slice:
\vspace{-1mm}
\begin{equation}
P(W,d,t) = \prod_{n=1}^{M} P(w_n | z_n , \beta_{t}) \cdot P(z_n | \theta_{d,t}) \cdot P(\theta_{d,t} | \alpha).
\end{equation}

Model estimation has some drawbacks under these assumptions. Posterior inference (model estimation of parameters conditioned on observed variables) is intractable due to the non conjugacy of Gaussians and multinomial distributions. Blei and Lafferty explored variational methods for posterior inference, discarding stochastic simulation (e.g. Gibbs sampling) due to computational difficulties inherent in the non conjugacy of Gaussians.  
To retain the sequential structure of topics over time, DTM fits a dynamic model with Gaussian variational observations ($\hat{\beta}_{k,1}, \ldots, \hat{\beta}_{k,t}, \ldots , \hat{\beta}_{k,T}$), fitting these parameters to minimize the Kullback-Leibler divergence between the resulting posterior and the true posterior. To mimic Gaussian variational observations, DTM uses Kalman filtering, which enables the use of backward-forward calculations in a linear state space model. Analogously, topic proportions $\theta_{t,d}$ are conditioned on free Dirichlet parameters $\gamma_{t,d}$ and topic indexes $z_{t,d,n}$ are conditioned on free multinomial parameters $\phi_{t,d,n}$: 
\vspace{-1mm}
\begin{eqnarray}
q(\beta_{k,1}, \ldots , \beta_{k,T} | \hat{\beta}_{k,1}, \ldots, \hat{\beta}_{k,T}), \hspace{8mm} \textmd{(Kalman parameters)} \\
q(\theta_{t,d} | \gamma_{t,d}), \hspace{6mm} \textmd{(Dirichlet parameters)} \\
q(z_{t,d,n} | \phi_{t,d,n}), \hspace{2mm} \textmd{(Multinomial parameters)}
\end{eqnarray}

Forward-backward calculations on Kalman parameters allows to estimate posterior mean and variance parameters ($m_t$ and $V_t$) in terms of Gaussian parameters ($\sigma^2$) over topics. As time sliced topics $\beta_t$ are conditioned on the immediate past time sliced topic $\beta_{t-1}$, and both are related by a Gaussian of $\sigma$ parameter, the variational state space model $\hat{\beta}_t$ is conditioned on $\beta_t$ and both are related by a Gaussian of $\hat{v}_t$ parameter. Then, in forward calculations, posterior mean and variance parameters ($m_t$ and $V_t$) are calculated from $\sigma$ and from the variational parameters $\hat{\beta_t}$ and $\hat{\sigma}_t$. In backward calculations the marginal mean $\widetilde{m}_{t-1}$ and variance $\widetilde{V}_{t-1}$ of $\beta_{t-1}$ depends on posterior mean $m_{t-1}$ and variance $V_{t-1}$, $\sigma$ and the one-step ahead marginal mean $\widetilde{m}_t$ and variance $\widetilde{V}_t$. Forward calculations use initial conditions $m_0$ and $V_0$ and backward calculations use initial conditions $\widetilde{m}_T = m_T$ and $\widetilde{V}_T = V_T$. The rest of the parameters are estimated using variational expectation maximization (VEM) as was proposed in the original posterior inference algorithm of LDA \cite{blei:03}.   

\subsection{Variational Inference Algorithm}

To estimate model parameters, DTM works using VEM and Variational Kalman filtering in a tandem. 
The inference algorithm starts initializing Kalman parameters using the LDA static VEM inference algorithm over the whole corpus, discarding timestamps. As an output of this process DTM obtains Kalman variational parameters ($\hat{\beta}_i$). Forward calculations are conducted to estimate posterior means and variances $m_t = \mathbb{E}(\beta_t | \hat{\beta}_{1:t})$ and $V_t = \mathbb{E}((\beta_t - m_ṭ)^2 | \hat{\beta}_{1:t})$ with initial conditions $m_0 = 0$ and $V_0 = \sigma^2 \cdot \textsc{e+03}$. Backward recurrences are used to estimate marginal means and variances $\widetilde{m}_{t-1} = \mathbb{E}(\beta_{t-1} | \hat{\beta}_{1:T})$ and $\widetilde{V}_{t-1} = \mathbb{E}((\beta_{t-1}-\widetilde{m}_{t-1})^2 | \hat{\beta}_{1:T})$ with initial conditions $\widetilde{m}_T = m_T$ and $\widetilde{V}_T = V_T$. Likelihood bound variables $\zeta_t = \sum_w e^{\widetilde{m}_{tw} + 0.5 \cdot \widetilde{V}_{tw}}$ are calculated for each topic in each time slice and $\beta_{t,k,n}$ for each topic, term and time slice in the corpus. 

After the initialization step, the inference algorithm runs the EM algorithm. The E-step uses the static LDA VEM inference algorithm, at document level, in chronological order according to document timestamps. Then, for each document a free Dirichlet parameter $\gamma_{t,d}$ is obtained, and for each word in each document a multinomial parameter $\phi_{t,d,n}$ is obtained. The E-step iterates until convergence following the LDA VEM convergence criteria. Then, the M-step runs bounding topic likelihoods. The process repeats the steps considered in the initialization process conditioned on $\phi_{t,d,n}$ model parameters. Forward and backward calculations are conducted reestimating variational Kalman parameters iterating until convergence following a topic likelihood criteria. At global level, E-step and M-step alternates until convergence, following a criteria that combines document and topic likelihoods.  

\subsection{Incremental learning on DTM}

A key aspect of Viscovery relies on incremental learning. As topics are used as opinion containers, the need to incorporate new opinions in a daily basis is a key aspect to keep information updated. To avoid the recalculation of the entire model, we extended DTM to allow incremental learning, updating the model to be consistent with new opinions but avoiding the recalculation of model parameters that depends on previous time slices. 

When a new document batch is aggregated into Viscovery, a set of unseen words may appears. Suppose that $Q$ new words are appended by the new batch to the model and assume that the batch size (number of documents in the batch) is $R$. Let $\mathcal{W}^{\textsc{new}} = \{ w_{M+1} , \ldots , w_{M+Q} \}$ be the set of new words and let $\mathcal{D}^{\textsc{new}} = \{ d_{N+1} , \ldots , d_{N+R} \}$ be the new batch. We need to aggregate to the model a set of new parameters. A first set of parameters is in dependence of $\mathcal{W}^{\textsc{new}}$ and previous slices $1:T$. Topic parameters included into the model are $\beta_{M+1,1:T}, \ldots , \beta_{M+Q,1:T}$. As new words were unobserved on previous slices, we set these parameters using $\beta_{\textsc{long-tail}}$, the value assigned by DTM to words in the long-tail of the model. In practice, by choosing at random any word in the tail of any topic, $\beta_{\textsc{long-tail}}$ achieves only small fluctuations (order $10^{-12}$). Analogously, we set $\hat{\beta}_{M+1,1:T}, \ldots , \hat{\beta}_{M+Q,1:T}$ as 
$\hat{\beta}_{\textsc{long-tail}}$, the value assigned by DTM to words in the long-tail of the Kalman variational parameters. Then we fit a static LDA over the new batch to obtain initial values for $\beta_{1:Q,T+1}$ parameters (model parameters for the new batch over the whole vocabulary). Mean and variance parameters (variational and marginal) are calculated using the forward-backward procedure at one step (one step ahead for forward calculation and one step behind for backward recursion). To avoid unnecessary computation costs, we discarded the recalculation of the entire chain of Kalman variational parameters, constraining inference only to dependencies between batches in slices $T$ and $T+1$. The constrained forward-backward calculation produces estimations for mean and variance in the new batch, and values for likelihood bounds $\zeta_{T+1}$ and $\beta_{T+1,k,n}$. 

Now we follow the EM procedure. 
Log likelihoods of topics and documents modeled in previous batches are retrieved to be included in the global likelihood criterion function used in the EM procedure. 
The E-step is conducted over the documents included in $\mathcal{D}^{\textsc{new}}$, obtaining estimates for $\gamma_{T+1,d}$ and $\phi_{T+1,d,n}$, $d \in \mathcal{D}^{\textsc{new}}$, likelihood bounds for each document. The E-step runs until convergence following the LDA VEM convergence criteria. The M-step runs bounding topic likelihoods. The process repeats the following cycle at word level for each document in the new batch: $\textsc{repeat:}$ $\textsc{topic bound est}$ $\rightarrow$ $\textsc{batch model updating}$ $\rightarrow$ $\textsc{new topic bound}$ $\rightarrow$ $\textsc{check convergence}$. As the M-step runs over the last chain of DTM, the convergence is very fast and the overall convergence is also very fast. In the appendix we give more details about how our incremental algorithm guarantees that the log likelihood $\log p(d_{1:T})$ is bounded from below using the Jensen's inequality.  

\section{Browsable Sentiment Analysis}
\label{browse_sent}

In this section we indicate how we produce a browsable sentiment analysis view of the data in Viscovery. 
Sentiment analysis is a key aspect of opinion mining tools and in Viscovery is a salient aspect that helps users to distinguish between subjective/neutral information. As a base service, the Novaviz API uses VADER \cite{vader:14} for sentiment sentence tagging. VADER provides three sentiment scores at sentence level: positive ($\textsf{sc}_{s}(\oplus)$), negative ($\textsc{sc}_{s}(\ominus)$) and neutral ($\textsc{sc}_{s}(\odot)$) scores, where $\textsf{sc}_{s}(\oplus) + \textsf{sc}_{s}(\ominus) + \textsf{sc}_{s}(\odot) = 1$. We recover for each sentence in our data these scores. 

\paragraph{Document level} A first level of aggregation considered in Viscovery is the document level. As opinions can be compounded by a number of sentences, sentiment scores need to be aggregated at opinion level. Let $d$ be a document indexed in Viscovery, and $s \in d$ the sentences that compounds $d$, where $|d|$ is the number of sentences of $d$. Sentiment scores at document level are obtained from:
\vspace{-1mm}
\begin{equation}
\textsc{sc}_{d}(*) = \sum_{s \in d} \frac{\textsc{sc}_{s}(*)}{|d|}, \quad \textmd{with } * \in \{ \oplus, \ominus, \odot \}
\end{equation}

Note that $\textsc{sc}_{d}(\oplus) + \textsc{sc}_{d}(\ominus) + \textsc{sc}_{d}(\odot) = 1$, as expected.

\paragraph{Topic level} A second level of aggregation considered in Viscovery is the topic level. As opinions are aggregated into topics, sentiment scores need to be aggregated at topic level to indicate the level of polarity of each topic. Let $z$ be a LDA latent variable, and $P(d | z)$ the membership probability given by DTM and defined in Equation \ref{eq:2}. Note that $P(* | z) = \sum_{d \in \mathcal{D}} P(* | d) \cdot P(d | z)$. For simplicity, we denote $P(* | z)$ by $\textsc{sc}_{z}(*)$. Then, sentiment scores at topic level are obtained from:

\begin{equation}
\textsc{sc}_{z}(*) = C_z \cdot \sum_{d \in \mathcal{D}} \textsc{sc}_{d}(*) \cdot P(d | z), \quad \textmd{with } * \in \{ \oplus, \ominus, \odot \}
\end{equation}

\noindent where $C_z = \frac{1}{\sum_{*} \textsc{sc}_{z}(*)}$. Note that $\textsc{sc}_{z}(\oplus) + \textsc{sc}_{z}(\ominus) + \textsc{sc}_{z}(\odot) = 1$, as expected.

\paragraph{Term level} At a high level of granularity Viscovery browses terms. To use sentiment analysis at term level, we need to estimate $P(* | w)$, denoted for simplicity by $\textsc{sc}_{w}(*)$. As $\textsc{sc}_{w}(*)$ can be expanded over latent variables by $\sum_{z \in Z} \textsc{sc}_{z}(*) \cdot P(z | w)$, using the Bayes rule on $P(z | w)$ we obtain $\textsc{sc}_{w}(*)$ as:

\begin{equation}
\textsc{sc}_{w}(*) = \sum_{z \in Z} \textsc{sc}_{z}(*) \cdot \frac{P(w | z) \cdot P(z)}{P(w)}, \quad \textmd{with } * \in \{ \oplus, \ominus, \odot \}
\end{equation}

Note that $\textsc{sc}_{w}(\oplus) + \textsc{sc}_{w}(\ominus) + \textsc{sc}_{w}(\odot) = 1$, as expected.

\paragraph{Using topics as proxies} Viscovery allows to browse opinions using topics as proxies. When a topic is picked in Viscovery, the sentiment view of the data can be projected to documents or terms. To show sentiment scores conditioned on topics, we reuse the scores defined in equations 11-13. Sentiment scores at document level conditioned on topics are defined by:
\vspace{-1mm}
\begin{equation}
\textsc{sc}_d(* | z) = \left( \sum_{w \in \mathcal{W}} \textsc{sc}_w(*) \cdot P(w | z) \right) \cdot P(z | d), \quad \textmd{with } * \in \{ \oplus, \ominus, \odot \} 
\end{equation}

Analogously, sentiment scores at term level conditioned on topics are defined by:
\vspace{-1mm}
\begin{equation}
\textsc{sc}_w(* | z) = \left( \sum_{d \in \mathcal{D}} \textsc{sc}_d(*) \cdot P(d | z) \right) \cdot P(z | w), \quad \textmd{with } * \in \{ \oplus, \ominus, \odot \} 
\end{equation}

This simple way to aggregate scores from sentence sentiment scores allows us to use sentiment analysis on DTM. 

\begin{figure*}[ht!]
\begin{center}
\includegraphics[width=11cm]{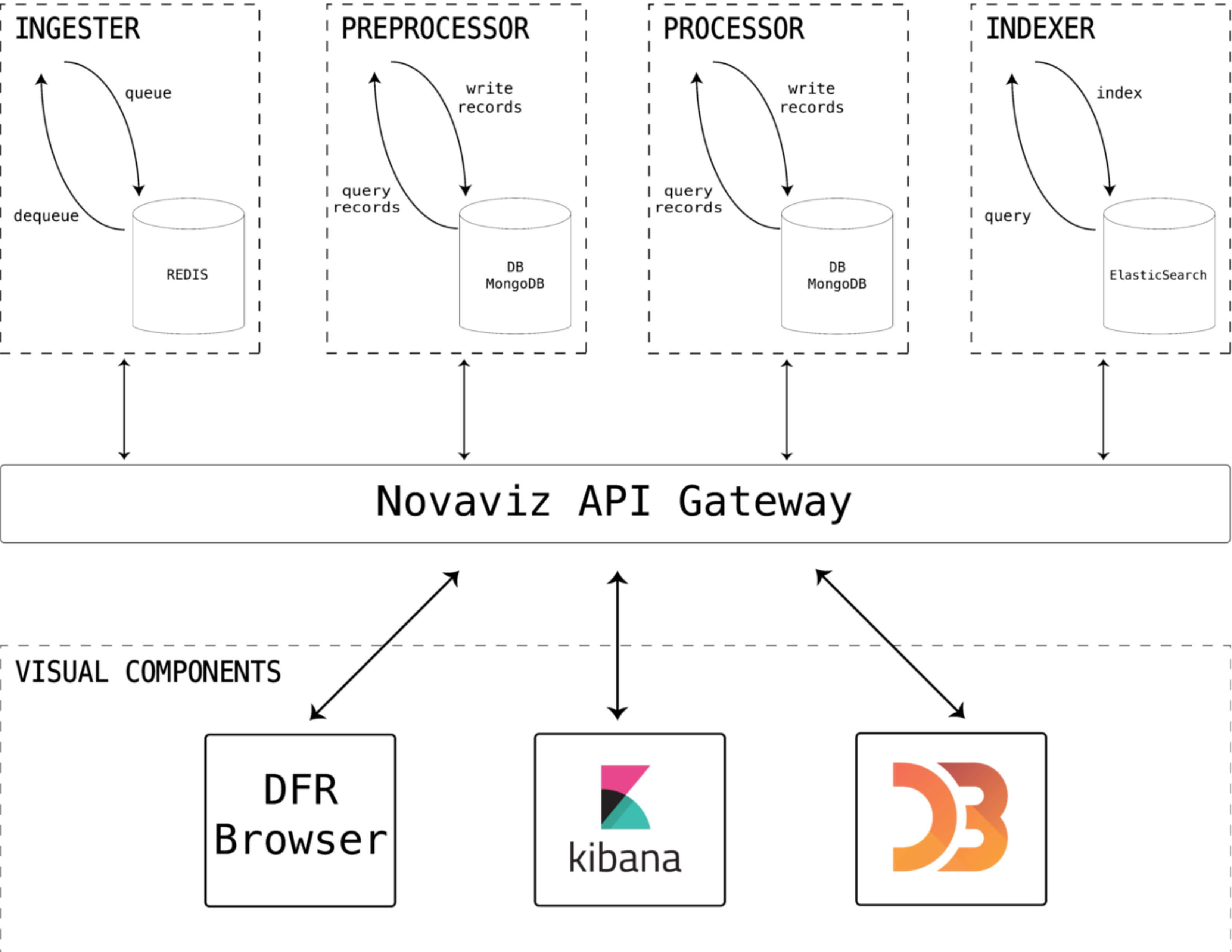}
\caption{Viscovery architectural diagram. Visualization components are connected to data processing components using the Novaviz API.}
\label{fig:1}
\end{center}
\end{figure*}

\section{Viscovery: Architecture and Design Principles}
\label{archi}

In this section we discuss how we integrate different algorithms to ingest opinions into Viscovery. We model different algorithms as micro services to develop a platform for trend tracking in opinion forums. A micro service architecture organizes the platform as a set of weakly coupled services where each service implements a set of encapsulated procedures. For example, a micro service in Viscovery corresponds to an indexer of opinionated tweets. Services in Viscovery are communicated using asynchronous protocols. We developed each service independently of the other. Indeed each micro service has its own database in order to be decoupled from other services.    

To develop Viscovery we create a start-up named Novaviz. 
The idea behind Novaviz is to develop tools for text data management. 
To accomplish this purpose we develop the Novaviz API Gateway, a list of services and functionalities implemented in Python, requested by four components: a) Data Ingester, b) Data Preprocessor, c) Data Processor, and d) Indexer. For visualization we use three libraries: a) DFR browser, b) Kibana, and c) D3. Visualization and processes are connected through micro services defined in Novaviz API Gateway, as is shown in the architectural diagram of Figure \ref{fig:1}.

\paragraph*{Data Ingester} This component is in charge of opinion recollection from heterogeneous sources as Twitter and web forums (e.g. http://www.reclamos.cl). It calls services from the Novaviz API. Among the services requested the most important is web scrapping, that allows Viscovery to retrieve opinions from web page forums. For storage, this component interacts with Redis~\footnote{https://redis.io/}, an open source (BSD licensed) in-memory data structure store, used as a cache database to support this process.

\paragraph*{Preprocessor} This component normalizes the text. It calls services from the Novaviz API such as stop words removal, caps normalization and punctuation removal. It allows Viscovery to create a vocabulary of keywords to describe opinions by content. For storage this component interacts with MongoDB~\footnote{https://www.mongodb.com/}, a noSQL database for document storage and retrieval.

\paragraph*{Processor} This component is in charge of text analysis and is the core component of Viscovery. It calls services from the Novaviz API as Dynamic Topic Models and Sentiment Analysis. For Dynamic Topic Models (DTM), the API wraps Gensim~\footnote{http://radimrehurek.com/gensim/}. Gensim is an implementation of topic modeling written in Python \cite{gensim}. It includes implementations of LDA, LSI and DTM. For sentiment analysis, the API wraps Vader~\footnote{https://github.com/cjhutto/vaderSentiment}, a rule-based model for sentiment analysis that uses a lexicon of English words \cite{vader:14}. As the preprocessor component, the processor interacts for storage with MongoDB, allowing to register each view of the data (e.g. topic model view) as a document view of each opinion, with the attributes leveraged by the respective view. For instance, from the sentiment view of an opinion, each document in MongoDB stores neutral, positive and negative scores at sentence level. Weights for topic membership are stored in the topic model view of each opinion. Then, documents in MongoDB will ingest the indexer, the component that provides data for opinion search and browsing. 

\paragraph*{Indexer} This component is in charge of opinion indexing. For each view of the data, we create an index allowing search and browsing at different levels of granularity. As opinions are clustered using topic models, browsing is conducted using topics as opinion aggregation containers. For each topic, each opinion register its membership score, which indicates the degree of membership of each opinion to the topic. As each topic is a probability distribution over words, we store the weights of each word per topic. As browsing is conducted over topics, the use of words to describe each topic is a key element of Viscovery. To integrate the sentiment view of the data, we index opinions and their related sentiment weights for search and browsing. To ingest these indexes, we recover the document views created by the processor in the previous step, processing and indexing them into Elasticsearch \footnote{https://www.elastic.co/}. Elasticsearch is a distributed, RESTful search and analytics engine capable of support searches over unstructured data implementing fast and efficient data access operations using inverted indexes. We use Elasticsearch indexes to support all the search and browsing operations in Viscovery. 

\paragraph*{DFR browser} To visualize opinion trends we started using DFR browser \footnote{https://agoldst.github.io/dfr-browser/}, a visualization tool that works over topic models to integrate data views into a single, coherent, and search\-able visualization of the data. As the code of DFR browser is available, we started working over DFR browser to cast this tool to our needs and requirements. DFR allows to search over topics, the basic search-able element in the visualization, and to disaggregate the information at topic level into documents and words by topic. 

\paragraph*{Kibana} Kibana is part of the suite provided by Elastic, named The Open Source Elastic Stack. The purpose of Kibana is to hand Elaticsearch visualizations.  

\paragraph*{D3.js} Another tool that we use for data visualizations is D3.js \footnote{https://d3js.org/}. D3.js is a JavaScript library for data visualizations compatible with HTML, SVG, and CSS. D3 follows a data-driven approach for data manipulation, using DOM as a standard for document representation. 

\section{Implementation Issues}
\label{imp_issues}

\subsection{Novaviz API}

The Novaviz API includes a list of services and functionalities. 
As the architecture of Viscovery is micro service oriented, the Novaviz API contains a list of reusable and generic services. Our API includes seven services:

\paragraph{Scrapper} This service extracts and recovers data from heterogeneous data sources as Twitter or opinion web forums (e.g. reclamos.cl). In the case of Twitter, it takes as seed a hashtag using the public API and producing a .json file compounded by the list of tweets that contains the hashtag. In the case of reclamos.cl we scrap the html source code of the forum recovering a semi-structured view of the forum in a .json file. The attributes included in the file are creation date, the complaint content (unstructured), the url (a permalink created by reclamos.cl for each opinion), and the title of the complaint. It is implemented using Scrapy ~\footnote{https://scrapy.org/}, a scrapper implemented in Python. The scrapper is called by the ingester component of Viscovery.

\paragraph{Corpus constructor} It takes scrapper outputs in .json format preprocessing the content to normalize the text. It starts tokenizing the text. Is in this service that caps, stop words, accents, punctuation and symbols are processed. We include a rule-based word removal by frequency. By default, words with one occurrence in the data are removed. In addition, a second rule-based word removal is included, removing words by length. Words with less than two chars are removed from the vocabulary. The constructor is language-flagged. Novaviz considers two languages, English and Spanish. By default, the constructor is set to English. The stopword list is customizable. In addition, the basis for time slicing can be defined here using a parameter with values in $\{ \textsc{daily}, \textsc{weekly}, \textsc{monthly}, \textsc{yearly} \}$. Output files produced by the corpus creator are \textit{foo}.dict (dictionary), \textit{foo}.lda-c (a row oriented file with one doc per row and entries indicating word occurrences), sliced.json (docIDs and timestamps). These files are used for LDA model fitting. The corpus constructor is called by the preprocessor component of Viscovery.      

\paragraph{LDA fitting} It takes \textit{foo}.dict (dictionary) and \textit{foo}.lda-c for LDA model fitting. It needs the number of topics as a parameter (five topics by default). LDA fitting wraps the Gensim implementation of LDA that is based on Gibbs sampling \cite{griffiths:04}. Output files produced by LDA fitting are stored in a directory that contains topic-word.json (truncated to the top-30 words per topic), doc-topic.json (topic proportions), frequency.txt (a list of words with their occurrences of the corpus), and \textit{foo} which corresponds to the model file, a coded view of the fitted model. LDA fitting is called by the processor component of Viscovery.

\paragraph{DTM fitting} Analogously to LDA fitting, this service wraps the Gensim implementation of DTM. It takes \textit{foo}.dictionary, \textit{foo}.lda-c and sliced.json for DTM fitting. Output files produced by DTM are topic-word.json with timestamps (one timestamp per time slice for each word in the dictionary), doc-topic.json (topic proportions), frequency.txt, and \textit{foo} which corresponds to the coded view of the DTM model. DTM fitting is called by the processor component of Viscovery. 

\paragraph{Sentiment Analysis} It takes the .json file produced by the Scrapper and conducts sentiment analysis using VADER. It works at three different levels of granularity, sentences, documents and topics. Output files produced by Sentiment Analysis are stored in .json files (with pairwise entries ID-sentiment score). A detailed discussion about how sentiment scores are calculated at different levels of granularity is provided in Section \ref{browse_sent}.

\paragraph{NER} It takes the .json file produced by the Scrapper service, and conducts named entity recognition using NLTK \footnote{http://www.nltk.org/} (Natural Language ToolKit, a Python implementation of NLP basic tools). Output files produced by NER are stored in .json files (with pairwise entries wordID-NER tag). 

\paragraph{POS} It takes the .json file produced by the Scrapper service, and conducts part-of-speech using NLTK. Output files produced by POS are stored in .json files (with pairwise entries wordID-POS tag). 

\paragraph{Topic scaling} It takes as input the \textit{foo} model file retrieved from LDA fitting. In addition it takes the \textit{foo}.dict (dictionary) and \textit{foo}.lda-c from corpus constructor to recover document lengths. Topic scaling wraps Principal Coordinate Analysis (PCoA) using the implementation provided in Scikit-bio \footnote{http://scikit-bio.org/}. Dimensionality reduction is conducted using PCoA towards a 2D embedding based on the Jensen-Shannon divergence between topics. The output file produced by topic scaling is a .json file with paiwise entries topicID - $\langle$x, y$\rangle$. Topic scaling is called by DFR browser for topic visualization.

\subsection{Elastic Indexes}

For data storage we use Elastisearch indexes. 
Elasticsearch provides services for data indexing and retrieval. 
Elastic indexes are key elements for opinion browsing, allowing to browse opinions at different levels of granularity. 
For each level of granularity we created a specific index in Elastic:

\paragraph{Opinions index} This index retrieves opinions using a docID as a search key. Each opinion is indexed at full content, allowing fast retrieval of opinions when documents are picked in Viscovery.

\paragraph{Topic-word index} This index retrieves top-K words per topic, using  topicID as a search key. For each word the index stores the membership level for the topic. 

\paragraph{Topic-document index} This index retrieves top-k documents per topic, using topicID as a search key. For each document the index stores the membership level for the topic.

\paragraph{Term frequency index} This index retrieves the frequency of each term using termID as a search key. 

\paragraph{Sentiment-document index} This index retrieves sentiment scores at document level using docID as a search key.

\paragraph{Sentiment-topic index} This index retrieves sentiment scores at topic level using topicID as a search key.

\paragraph{Sentiment-sentence index} This index retrieves sentiment scores at sentence level using sentenceID as a search key.

\subsection{DFR cast}

DFR (Data For Research)~\footnote{http://agoldst.github.io/dfr-browser/demo} is a visualization tool that produces global data views avoiding unnecessary accesses to documents at finer granularity levels. 
It produces a global first data view comprising document contents using topics as containers and words as topic descriptors. 
Topics can be picked in the global data view showing the most relevant words of the selected topic as a list. 
The temporal evolution of the topic is showed in the topic view. Lengthwise selection on the timeline of the topic exhibits data responsiveness, updating top-documents at topic level. 
Along with top document at topic level, a list of top-50 words is showed.

\begin{figure}[ht!]
\begin{center}
\includegraphics[width=8cm]{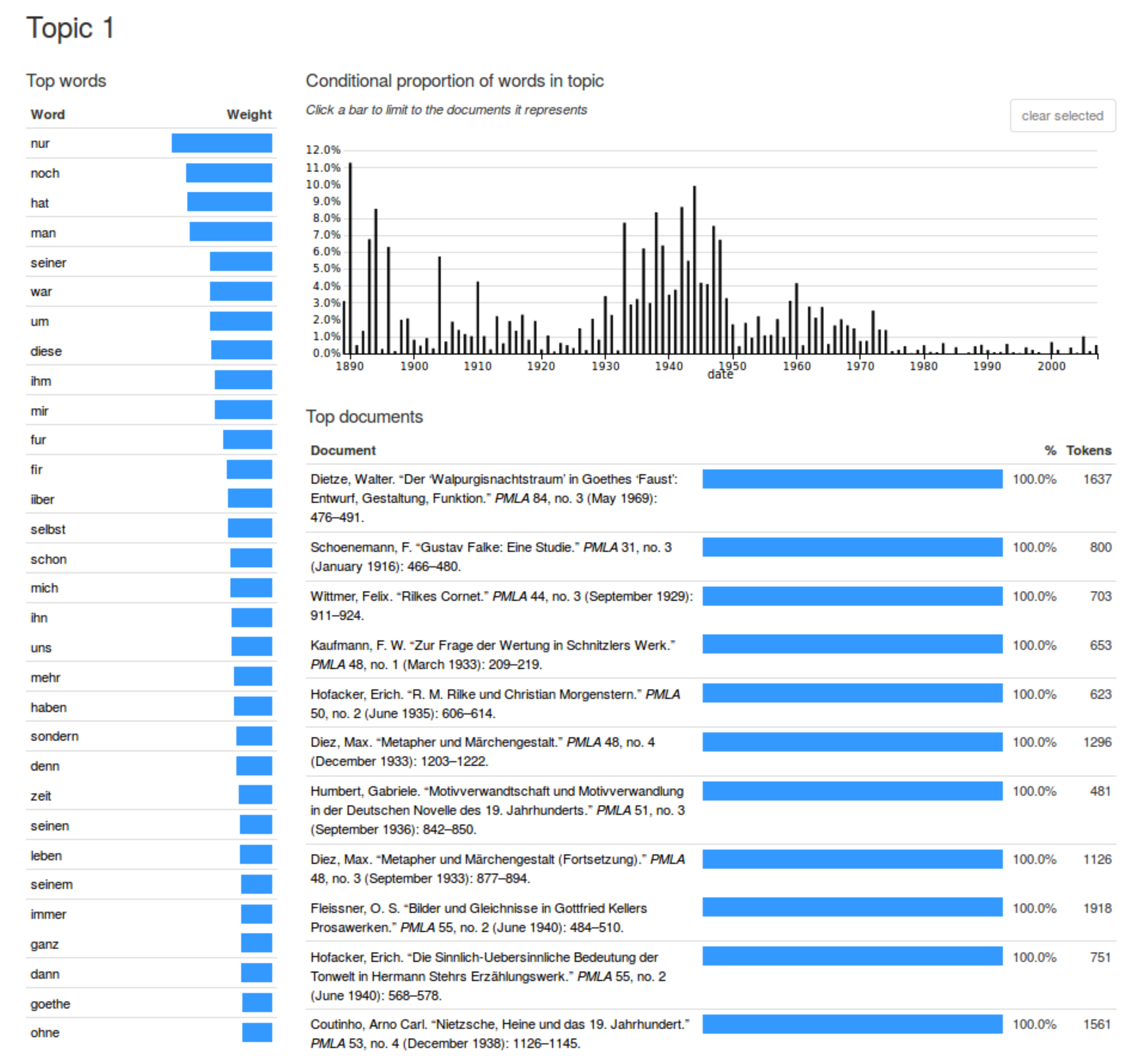}
\caption{DFR topic-view.}
\label{fig:2}
\end{center}
\end{figure}

The document list includes three attributes per document: the title, the degree of membership of the document to the selected document, and the number of tokens that compound it. 
These lists include the top-20 most salient documents per topic in terms of degree of membership. The topic view provided by DFR is shown in Figure \ref{fig:2}.

\begin{figure}[ht!]
\begin{center}
\includegraphics[width=8cm]{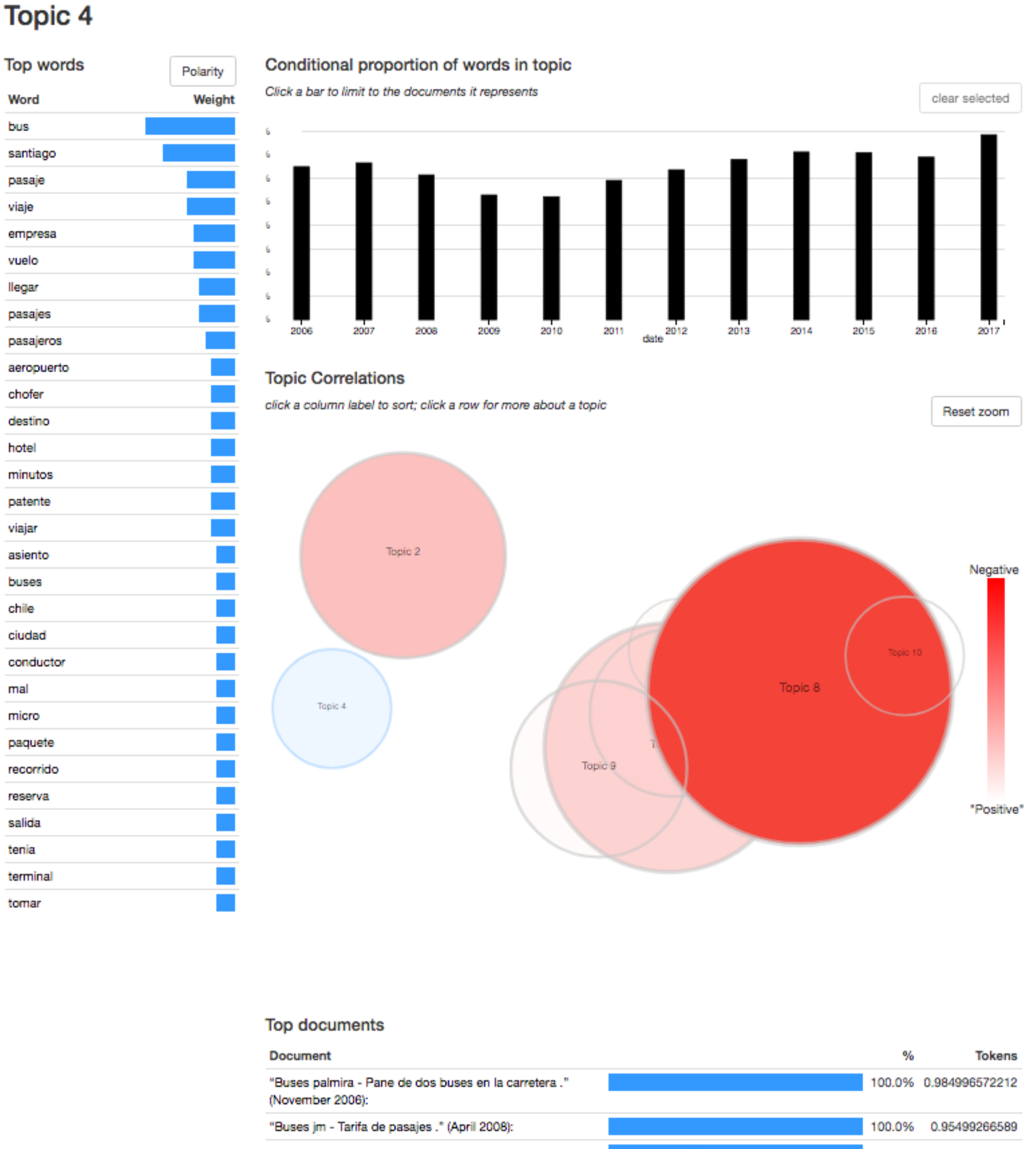}
\caption{Sentiment DFR topic-view.}
\label{fig:3}
\end{center}
\end{figure}

We extended the DFR topic view to include sentiment analysis and a 2D topic embedding global view between the timeline and the list of top documents. 
To en-chase the topic embedding we modified the DRF topic view. By default, DFR does not include sentiment visualizations. Then two files, sentiment scores at term and document levels were included to allow sentiment visualizations. These files were used to indicate the polarity of topics, documents and words. 
At topic level, each topic was colored according to its polarity, using a white/red color palette (negative scores were represented in red). In addition we included a button in the top words list to change the length of each word bar according to objective/subjective scores. A screen shoot of our sentiment DFR extension is shown in Figure \ref{fig:3}.

\section{Data slices and preliminary results}
\label{views}

\subsection*{Incremental learning testing}

We evaluate our incremental version of DTM to measure speed up and model quality in terms of topic coherence \cite{wallach:09}. We expect to reduce the computational time involved in model fitting avoiding a retrogress in terms of topic coherence. To test this aspect of our proposal, we run ten trials of model fitting for a corpus, with and without incremental learning on the last time slice. We used as test data a curated dataset to evaluate topic coherence provided by Greene \& Cross \cite{Greene:16} that comprises news about the political agenda of the European Parliament. The dataset is divided into four time slices and is compounded by 1324 news articles classified into a manually-specified number of topics, helping to evaluate topic coherence. Mean and variance of coherence and average computational times involved in both algorithms are reported in Table \ref{tab:coherence}.

\begin{table}[h!]
\begin{center}
\begin{tabular}{l|ccc}
Algorithm         &   Mean Coh.   &  Var Coh.    &   Avg Comp. Time   \\
\hline
DTM               &      -1.5747  &   0.0047     &   2:06:56           \\ 
DTM + Seq. Upd.   &      -1.5373  &   0.0063     &   1:47:37 + 0:11:50 \\
\end{tabular}
\end{center}
\caption{Topic coherence and computational times in DTM and DTM+seq update. Mean and variance over ten runs per algorithm are reported.}
\label{tab:coherence}
\end{table}
\vspace{-1mm}
As expected, the mean computational time involved in DTM + Seq. Upd. is less than the time registered by DTM, with only 11 minutes spent in the fourth slice. This result indicate that the most expensive step of the algorithm is the Kalman variational inference and as our proposal constraints this step two the last slice, it reduces the cost involved. As the data used for this experiment is small (we used this dataset to estimate topic coherence) the difference between both algorithms in terms computational time is small. In Reclamos.cl, a big data set with more than 200,000 complains that we indexed in Viscovery, the difference between both algorithms is high. If we use DTM over the whole dataset, model fitting takes 14.7 hours. On the other hand, using sequential update over the last time slice it takes 1.91 hours. Note that without sequential update, we will need to retrain the whole model for each new slice and our proposal avoids this with a speed up of almost 8x. Surprisingly the time reduction does not affect the quality of the model in terms of topic coherence as is shown in Table \ref{tab:coherence}. In fact, our proposal achieves a slight improvement over DTM at a cost of a higher variance between the different trials.  

\subsection{Data slices over Reclamos.cl}

We are developing Viscovery implementing new functionalities. In fact, the current version of Viscovery implements browsable sentiment analysis at topic and word levels. Currently we are working on the implementation of sentiment analysis at document level, according to the proposal introduced in Section \ref{browse_sent}. Viscovery allows to browse opinions using topics as proxies of opinions. We indexed into Viscovery 12 years of data from Reclamos.cl, a Chilean forum for complaints abouts companies, marks and institutions. The dataset contains 201,969 different complaints. Retail, government, banks and universities are among the most frequent subjects of opinions. 

\begin{figure*}[h!]
\begin{center}
\includegraphics[width=12.5cm]{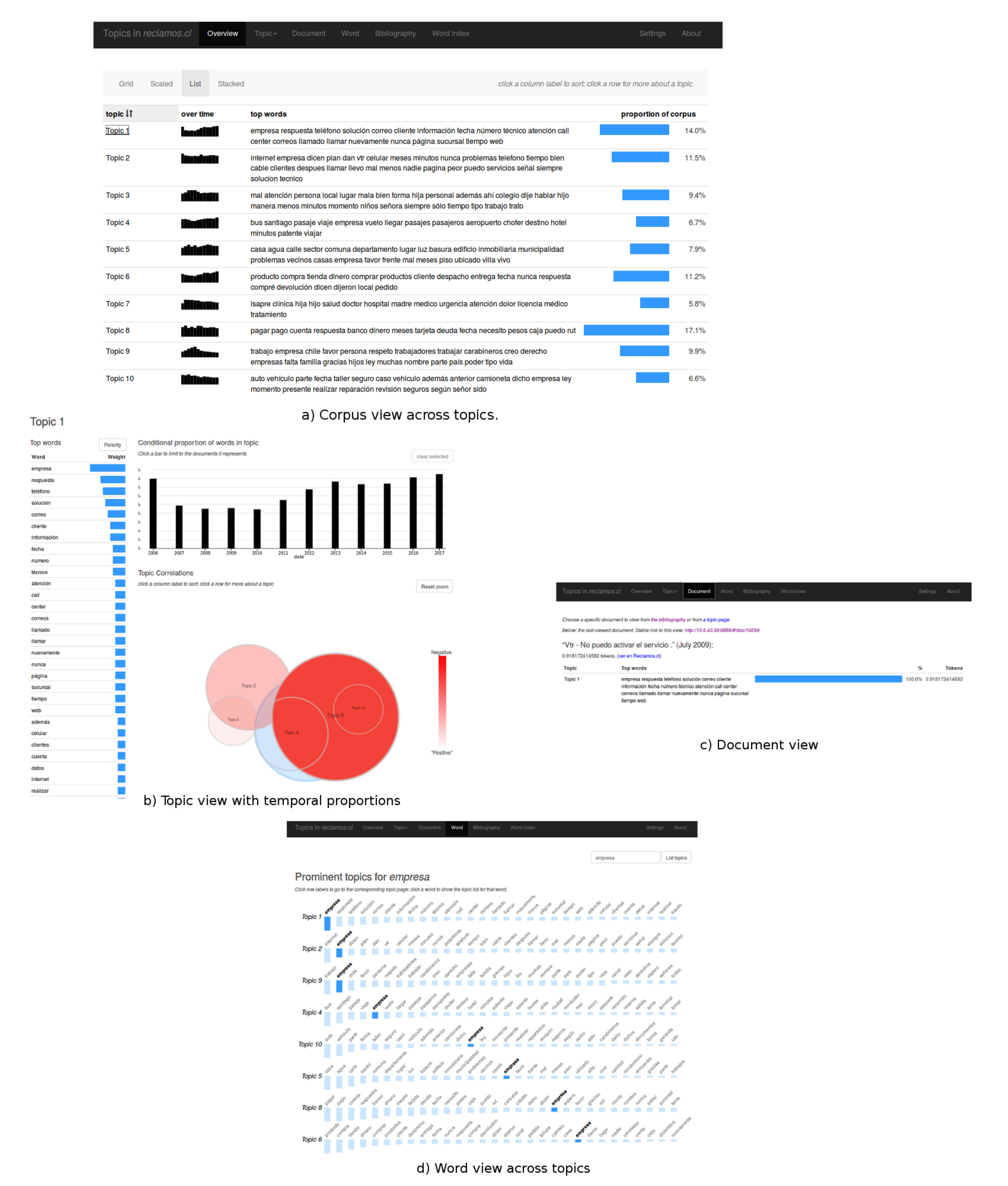}
\caption{Viscovery data views. Three data slices are deployed from the a) Corpus view (list of topics and temporal proportions) after topic selection: b) Topic view, which includes top words per topic, topic proportions on time and topic embedding, c) Document view (membership of the document to the given topic), and d) Word view across topics, showing the ranking of the word in each topic where the word is prominent. }
\label{fig:4}
\end{center}
\end{figure*}

The size of the vocabulary after stopword removal is 86,723 terms. 
We used 18 hours to create the corpus of reclamos using Viscovery. Reclamos.cl is a very active site in Chile, reporting an overall of 90,128 persons contacted by companies after complaint publication (a significant number in proportion to the Chilean population). 

Running DTM using years as timestamps, we achieved 12 time slices of the data. We used the default value for the number of topics, set as 10 for this example. Data slices for this corpus are shown in Figure \ref{fig:4}. 
As Figure \ref{fig:4} shows, the a) Corpus view (overview) has four alternatives for corpus deployment: grid, scaled, list and stacked. We show topics using lists as data view. The list of topics include topics proportions over time, top words per topic and the proportion of the topic in the corpus. When a topic is selected (we click topic 1 for this example), the b) Topic view is deployed. A topic view shows the list of top words for the topic, sorted in decreasing order according to the proportion on the topic. If the polarity bottom is pressed the bars are modified according to the sentiment weight of the word in the topic. For this version of Viscovery, the bar size is proportional to the sum of positive and negative scores. Currently we are implementing an extension that produces two bars per term, one per polarity. Topic embeddings are also shown in this view, illustrating the correlation of topic (distances in the topic embedding). The polarity orientation of each topic is shown using a color bar, where negative-biased topics are indicated with red shades. A list of top-documents per topic is shown below the embedding (omitted in this figure) and the user can select a specific opinion. In this case, Viscovery deploys the c) Document view, where the top words of the document in the given topic are shown. The subject and the date of the complaint is shown at the top of the view. Finally, Viscovery provides a d) Term view across topics, showing how relevant is a given word across topics. In the example, we show the word view using the term 'company', and as expected, this word is used in many topics of reclamos.cl, with different levels of membership. 

\section{Conclusions}
\label{conc}

We present Viscovery, a tool for opinion browsing and trend tracking. Key elements of Viscovery are Dynamic Topic Models (DTM) and our extension of DTM for sequential updating. We include sentiment analysis in Viscovery starting from sentiment scores at sentence level and then, conducting aggregation across topics and documents. This approach is simple and effective. For visualization we use DFR browser, extending DFR to include topic embeddings and sentiment analysis.

Currently we are extending Viscovery to include more functionalities. 
Among these functionalities we are working on the sentiment-document view, topic evolution tracking view and opinion search module. We are implementing these modules using Kibana and D3, two visual components considered in Viscovery not included in the current version. In addition, we are using Viscovery to index more sources, as opinions retrieved from Twitter and Reddit. 

\section*{Acknowledgment}

This work was supported by the Fondef VIU 15E0085 project of the National Agency of Science and Technology, Conicyt, Chile. 

\bibliographystyle{ACM-Reference-Format}
\bibliography{sigproc} 

\section*{Appendix. Lower bound of the likelihood for the incremental algorithm}

In this section we give details of the incremental algorithm that maximizes the lower bound of the likelihood on $\log p(d_{1:T})$. 
This section is an extension of the appendix provided in \citet{blei:06} where the lower bound is calculated for the static algorithm. For the incremental version of the inference algorithm, we only need to calculate the terms for the last time slice $T$. The first term of the lower bound is:
\vspace{-1mm}
\begin{equation}
\begin{array}{l}
\mathbf{E}_{q} \log_{p}(\beta_T | \beta_{T-1}) = -\frac{V}{2}(\log\sigma^2 + \log(2\sigma^2) \\ \\-  \frac{1}{2\sigma^2}\mathbf{E}_{q}(\beta_T - \beta_{T-1})^{T}(\beta_T - \beta_{T-1}) \\
\\ = - \frac{V}{2}(\log\sigma^2 + \log 2\pi) - \frac{1}{2\sigma}|| \tilde{m_T} - \tilde{m_{T-1}}||^2 \\
\\ - \frac{1}{\sigma^2} Tr(\tilde{V_T}) + \frac{1}{2\sigma^2}(Tr(\tilde{V_0}) - Tr(\tilde{V_T}))
\end{array}
\end{equation}

The second term is:
\begin{equation}
\begin{array}{l}
\mathbf{E}_{q} \log_p(d_{T} | \beta_{T}) = \\
\\\ \quad  \sum_{w}n_{tw}\mathbf{E}_{q}(\beta_{w} - \log \sum_{w}\exp{(\beta_{w})})
\\ \\ \quad \geq  \sum_{w}n_{w}\tilde{m}_{w} - n_{w}\zeta_{T}^{-1}\sum_w{\exp(\tilde{m_T} + \tilde{V}_{w}/2)} \\
\\ \quad  \qquad  + n_{T} - n_{T}\log{\zeta^{-1}_{T}}
\end{array}
\end{equation}

where $n_{T} =\sum_{w}n_{w}$. The third term is the entropy $H(q)  = \frac{1}{2}\sum_{w} \log \tilde{V}_{w} + \frac{V}{2}\log 2 \pi$. 
The term $\frac{V}{2}\log 2 \pi$ is canceled in term 1 and the entropy. In term 2 
\begin{equation}
\begin{array}{l}
n_{T}\zeta_{T}^{-1}\sum_w{\exp(\tilde{m_T} + \tilde{V}_{w}/2)} = n_{T}\zeta_{T}^{-1}\zeta_{T} = n_{T} \\
\end{array}
\end{equation}

The new term $ - n_{T}$ is canceled with the corresponding $n_{T}$. Then, the bound can be obtained as

\begin{equation}
\begin{array}{l}
 = - \frac{V}{2}(\log\sigma^2) - \frac{1}{2\sigma}|| \tilde{m_T} - \tilde{m_{T-1}}||^2  - \frac{1}{\sigma^2} Tr(\tilde{V_T}) \\ \\ 
 + \frac{1}{2\sigma^2}(Tr(\tilde{V_0}) - Tr(\tilde{V}_T)) + \sum_{w}n_{w}\tilde{m}_{w} - n_{T}\log{\tilde{\zeta_{T}}} \\ \\
+ \frac{1}{2}\sum_{w} \log \tilde{V}_{w} 
\end{array}
\end{equation}

\end{document}